\documentclass[referee]{jfm}

\input psfig

\title{Absence of slow transients, and the effect of imperfect vertical alignment, in turbulent Rayleigh-B\'enard convection}
\author{Guenter Ahlers, Eric Brown, and Alexei Nikolaenko}   

\affiliation{Department of Physics and iQUEST,\\ University of
California, Santa Barbara, CA  93106}
\begin{document}

\maketitle

\begin{abstract}
We report experimental results for the influence of a tilt angle $\beta$ relative to gravity on turbulent Rayleigh-B\'enard convection of cylindrical samples. The measurements were made at Rayleigh numbers $R$ up to $10^{11}$ with two samples of height $L$ equal to the diameter $D$ (aspect ratio $\Gamma \equiv D/L \simeq 1$), one with  $L \simeq 0.5$ m (the ``large" sample) and the other with $\L \simeq 0.25$ m (the ``medium" sample).  The fluid was water with a Prandtl number $\sigma = 4.38$. 

In contrast to the experiences reported by \cite{CRCC04} for a similar sample but with $\Gamma \simeq 0.5$ ($D = 0.5$ and $L = 1.0$ m),  we found no long relaxation times. 

For $R = 9.4\times 10^{10}$ we measured the Nusselt number $ \cal N $ as a function of $\beta$ and obtained a small $\beta$ dependence given by $\ {\cal N}(\beta) = {\cal N}_0  [1 - (3.1\pm 0.1)\times 10^{-2}|\beta|]$ when $\beta$ is in radian. This depression of $\cal N$ is about a factor of 50 smaller than the result found by  \cite{CRCC04} for their $\Gamma \simeq 0.5$ sample. 

We measured side-wall temperatures at eight equally spaced azimuthal locations on the horizontal mid-plane of the sample and used them to obtain cross-correlation functions between opposite azimuthal locations.  The correlation functions had Gaussian peaks centered about $t_1^{cc} > 0$ that corresponded to half a turn-over time of the large-scale circulation (LSC) and yielded Reynolds numbers $R_e^{cc}$ of the LSC. 
For the large sample and $R = 9.4\times 10^{10}$ we found $R_e^{cc}(\beta) = R_e^{cc}(0)\times [1 + (1.85\pm 0.21)  |\beta| - (5.9\pm 1.7)  \beta^2]$. Similar results were obtained also from the auto-correlation functions of individual thermometers. These results are consistent with measurements of the amplitude $\delta$ of the azimuthal  side-wall temperature-variation at the mid-plane that gave $\delta(\beta) = \delta(0)\times [1 + (1.84 \pm 0.45)  |\beta| - (3.1 \pm 3.9) \beta^2]$ for the same $R$. An important conclusion is that the increase of the speed (i.e. of $R_e$) with $\beta$ of the LSC does not significantly influence the heat transport. Thus the heat transport must be determined primarily by the instability mechanism operative in the boundary layers, rather than by  the rate at which ``plumes" are carried away by the LSC. This mechanism apparently is independent of $\beta$.

Over the range $10^9 \stackrel {<}{_\sim} R \stackrel {<}{_\sim} 10^{11}$ the enhancement of $R_e^{cc}$ at constant $\beta$ due to the tilt could be described by a power law of $R$ with an exponent of $-1/6$, consistent with a simple model that balances the additional buoyancy due to the tilt angle by the shear stress across the boundary layers.

Even a small tilt angle dramatically suppressed the azimuthal meandering  and the sudden reorientations characteristic of the LSC in a sample with $\beta = 0$. 

For large $R$ the azimuthal mean of the temperature at the horizontal mid-plane differed significantly form the average of the top- and bottom-plate temperatures due to non-Boussinesq effects, but within our resolution was independent of $\beta$.

\end{abstract}

\section{Introduction}
\label{intro}

Turbulent convection in a fluid heated from below, known as Rayleigh-B\'enard convection (RBC), has been under intense study for some time [for reviews, see {\it e.g.} \cite{Si94,Ka01,AGL02}].  A central prediction of models for this system [\cite{Kr62,CGHKLTWZZ89,SS90,GL01}] is the heat transported by the fluid. It is usually described in terms of the Nusselt number
\begin{equation}
{\cal N} = \frac{Q L}{A \lambda \Delta T}
\label{eq:nusselt}
\end{equation}
where $Q$ is the heat current, $L$ the cell height, $A$ the cross-sectional area, $\lambda$ the thermal conductivity, and $\Delta T$ the applied temperature difference.
The Nusselt number depends on the Rayleigh number
\begin{equation}
R = \alpha g \Delta T L^3/\kappa \nu
\label{eq:R}
\end{equation}
and on the Prandtl number 
\begin{equation}
\sigma = \nu/\kappa\ .
\end{equation}
Here $\alpha$ is the isobaric thermal expansion coefficient, $g$ the acceleration of gravity, $\kappa$ the thermal diffusivity, and $\nu$ the kinematic viscosity.

An important feature of turbulent RBC is the existence of a large-scale circulation (LSC) of the fluid [\cite{KH81}]. For cylindrical samples of aspect ratio $\Gamma \equiv L/D \simeq 1$ the LSC is known to consist of a single cell, with fluid rising along the wall at some azimuthal location $\theta$  and descending along the wall at a location $\theta + \pi$ [see, for instance, \cite{QT01a}]. As $\Gamma$ decreases, the nature of the LSC is believed to change. For $\Gamma \stackrel {<}{_\sim} 0.5$ it is expected [\cite{VC03,SV05,SXX05}] that the LSC consists of two or more convection cells, situated vertically one above the other. Regardless of the LSC structure, the heat transport in turbulent RBC is mediated by the emission of hot (cold) volumes of fluid known as ``plumes" from a more or less quiescent boundary layer above (below) the bottom (top) plate. These plumes are swept away laterally by the LSC and rise (fall) primarily near the side wall. Their buoyancy helps to sustain the LSC.

In a recent paper \cite{CRCC04} reported measurements using a cylindrical sample of water with $\sigma \simeq 2.33$ and with $L = 1$ m and $D = 0.5$ m for $R \simeq 10^{12}$. Their sample thus had an aspect ratio $\Gamma \simeq 0.5$ at the borderline between a single-cell and a multi-cell LSC. They found exceptionally long relaxation times of $\cal N$ that they attributed to a switching of the LSC structure between two states. Multi-stability was observed also in Nusselt-number measurements by \cite{RCCH04} for a $\Gamma = 0.5$ sample (see also \cite{NBFA05} for a discussion of these data). Chill\'a et al. also found that $\cal N$ was reduced by tilting the sample through an angle $\beta$ relative to gravity by an amount given approximately by  
${\cal N}(\beta)/ {\cal N}(0)  \simeq 1 - 2 \beta$
when $\beta$ is measured in radian. A reduction by two to five  percent of $\cal N$ (depending on $R$) due to a tilt by $\beta \simeq 0.035$ of a $\Gamma = 0.5$ sample was reported as well recently by \cite{SXX05}, although in that paper the $\beta$-dependence of this effect was not reported. Chill\'a et al. developed a simple model that yielded a depression of $\cal N$ for the two-cell structure that was consistent in size with their measurements. Their model also assumes that no depression of $\cal N$ should be found for a sample of aspect ratio near unity where the LSC is believed to consist of a single convection cell; they found some evidence to support this in the work of \cite{BTL95}. Indeed, recent measurements by \cite{NBFA05} for $\Gamma = 1$ gave the same $\cal N$ within 0.1 percent for a level sample and a sample tilted by 0.035 rad.

In this paper we report on a long-term study of RBC in a cylindrical sample with $\Gamma \simeq 1$. As expected, we found no long relaxation  times because the LSC is uniquely defined. The establishment of a statistically stationary state  after a large change of $R$ occurred remarkably quickly, within a couple of hours, and thereafter there were no further long-term drifts over periods of many days. 

We also studied the orientation $\theta_0$ of the circulation plane of the LSC by measuring the side-wall temperature at eight azimuthal locations [\cite{BNA05b}]. With the sample carefully leveled ({\it i.e.} $\beta = 0$) we found $\theta_0$ to change erratically, with large fluctuations. There were occasional relatively rapid reorientations, as observed before by \cite{SBN02}.  The reorientations usually consisted of relatively rapid rotations, and rarely were reversals involving the cessation of the LSC followed by its re-establishment with a new orientation. This LSC dynamics yielded a broad probability distribution-function $P(\theta_0)$, although a preferred orientation prevailed. When the sample was tilted relative to gravity through an angle $\beta$, a well defined new orientation of the LSC circulation plane was established, $P(\theta_0)$ became much more narrow, and virtually all meandering and reorientation of the LSC was suppressed.

We found that $\cal N$ was reduced very slightly by tilting the sample. We obtained $\ {\cal N}(\beta) = {\cal N}_0  [1 - (3.1\pm 0.1)\times 10^{-2}|\beta|]$. This effect is about a factor of 50 smaller than the one observed by Chill\'a et al. for their $\Gamma = 0.5$ sample.

From side-wall-temperature measurements at two opposite locations we determined time cross-correlation functions $C_{i,j}$. The $C_{i,j}$ had a peak that could be fitted well by a Gaussian function, centered about a characteristic time $t^{cc}_1$ that we interpreted as corresponding to the transit time needed by long-lived thermal disturbances to travel with the LSC from one side of the sample to the other, i.e. to half a turnover time of the LSC. We found that the $\beta$-dependence of the corresponding Reynolds number $R_e^{cc}$ is given by $R_e^{cc}(\beta) = R_e^{cc}(0)\times [1 + (1.85\pm 0.21)  |\beta| - (5.9\pm 1.7)  \beta^2]$. A similar result was obtained from the auto-correlation functions of individual thermometers. Thus there is an ${\cal O}(1)$ effect of $\beta$ on $R_e$, and yet the effect of $\beta$ on $\cal N$ was seen to be nearly two orders of magnitude smaller. We also determined the temperature amplitude $\delta$ of the azimuthal temperature variation at the mid-plane. We expect $\delta$ to be a monotonically increasing function of the speed of the LSC passing the mid-plane, i.e. of the Reynolds number.  We found  $\delta(\beta) = \delta(0)\times [1 + (1.84 \pm 0.45)  |\beta| - (3.1 \pm 3.9) \beta^2]$. Thus, for small $\beta$  its $\beta$-dependence is very similar to that of the Reynolds number.

From the large effect of $\beta$ on $R_e$ and the very small effect on $\cal N$ we come to the important conclusion that the heat transport in this system is not influenced significantly by the strength of the LSC. This heat transport thus must be determined primarily by the efficiency of instability mechanisms in the boundary layers. It seems reasonable that these mechanisms should be nearly independent of $\beta$ when $\beta$ is small. This result is consistent with prior measurements by \cite{CCL96}, who studied the LSC and the Nusselt number in a sample with a rectangular cross section. They inserted vertical grids above (below) the bottom (top) plate that suppressed the LSC, and found that within their resolution of a percent or so the heat transport was unaltered. Their shadowgraph visualizations beautifully illustrate that the plumes are swept along laterally by the LSC when there are no grids and rise or fall vertically due to their buoyancy in the presence of the grids. \cite{CCL96} also studied the effect of tilting their rectangular sample by an angle of 0.17 rad. Consistent with the very small effect of tilting on $\cal N$ found by us, they  found that within their resolution the heat transport remained unaltered.

We observed that the sudden reorientations of the LSC that are characteristic of the level sample are strongly suppressed by even a small tilt angle.

\begin{figure}
\centerline{\psfig{file=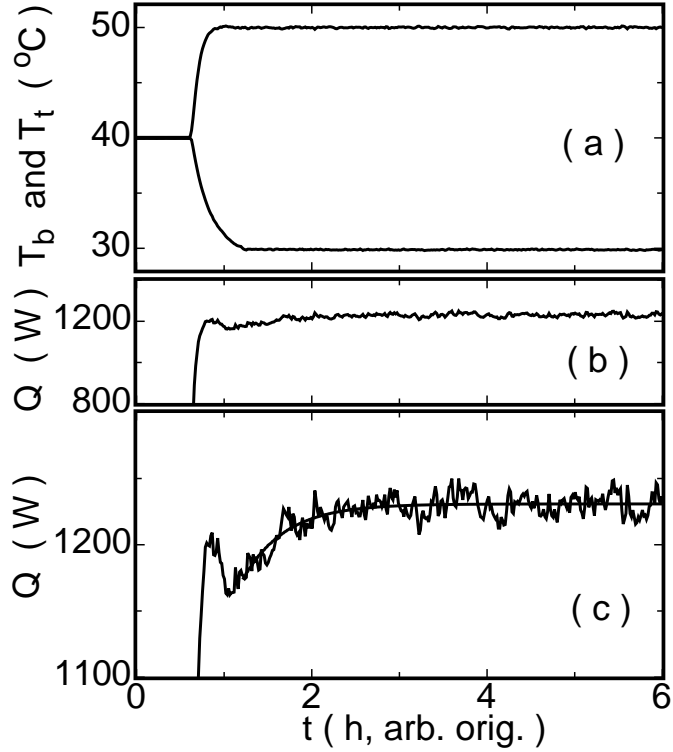,width=3.5in}}
\caption{Time evolution of the top- and bottom-plate temperatures and of the heat current for the large sample. Initially the temperature difference was $\Delta T \simeq 0$. At $ t = 0.6$ hours the top- and bottom-plate regulators were given new set points corresponding to $\Delta T = 20^\circ$C ($R = 9.43\times 10^{10}$). (a): Top- (lower data set, $T_t$) and bottom- (upper data set, $T_b$) plate temperatures. (b): Heat current $Q$ delivered by the temperature regulator designed to hold $T_b$ at a specified value. (c): Heat current $Q$ on an expanded scale. The solid line represents a fit of an exponential function to the data for $t > 1.2$ h that gave a relaxation time $\tau_Q = 0.48$ h.}
\label{fig:transients}
\end{figure}

\section{Apparatus and Data Analysis}
\label{sec:apparatus}

For the present work we used the ``large" and the ``medium" sample and apparatus described in detail by \cite{BNFA05a}. Copper top and bottom plates each contained five thermistors close to the copper-fluid interface. The bottom plate had imbedded in it a resistive heater capable of delivering up to 1.5 kW uniformly distributed over the plate. The top plate was cooled via temperature-controlled water circulating  in a double-spiral channel. For the Nusselt-number measurements a temperature set-point for a digital feedback regulator was specified. The regulator read one of the bottom-plate thermometers at time intervals of a few seconds and  provided appropriate power to the heater. The top-plate temperature was determined by the temperature-controlled cooling water from two Neslab RTE740 refrigerated circulators.

Each apparatus was mounted on a base plate that in turn was supported by three legs consisting of long threaded rods passing vertically through the plate. The entire apparatus thus could be tilted by an angle $\beta$ relative to the gravitational acceleration by turning one of the rods. The maximum tilt angle attainable was 0.12 (0.21) rad for the large (medium) sample.

The Nusselt number was calculated using the temperatures recorded in each plate and the power dissipated in the bottom-plate heater. The side wall was plexiglas of thickness 0.64 cm (0.32 cm) for the large (medium) sample. It determined the length $L$ of the sample. Around a circumference the height was uniformly $50.62 \pm 0.01$ cm ( $24.76 \pm 0.01$ cm) for the large (medium) sample. The inside diameter was $D = 49.70 \pm 0.01$ cm ( $D = 24.84 \pm 0.01$ cm) for the large (medium) sample. The end plates had anvils that  protruded into the side wall, thus guaranteeing a circular cross section near the ends. For the large sample we made measurements of the outside diameter near the half-height after many months of measurements and found that this diameter varied around the circumference by less than 0.1\%. 

Imbedded in the side wall and within 0.06 cm of the fluid-plexiglas interface were eight thermistors, equally spaced azimuthally and positioned vertically at half height of the sample. They yielded a relatively high (low) temperature reading at the angular positions where there was up-flow (down-flow) of the LSC. A fit of
\begin{equation}
T_i = T_c + \delta ~ cos(i \pi / 4 - \theta_0),\ i = 0, \ldots, 7
\label{eq:T_i}
\end{equation}
yielded the mean  center temperature $T_c$, the angular orientation $\theta_0$ of the LSC (relative to the location of thermistor 0), and a measure $\delta$ of the LSC strength. 

We expect the size of $\delta$ to be determined by the heat transport across a viscous boundary layer separating the LSC from the side wall. Thus  $\delta$ should be a monotonically increasing function of the LSC Reynolds number $R_e$ because the boundary-layer thickness  is expected to decrease with $R_e$ as $1/R_e^{1/2}$, and because the azimuthal temperature variation carried by the LSC near the boundary layer increases with $R$ and thus with $R_e$. However, the precise relationship between $\delta$ and $R_e$ is not obvious. Experimentally we find, over the range $5\times 10^9 <  R < 10^{11}$ and for the large sample, that $\delta$ is related to $R$ by an effective power law $\delta \propto R^{0.81}$, whereas $R_e \propto R^{0.50}$ in this range, yielding $\delta \propto R_e^{1.62}$. We would then expect that $\delta$ and $R_e$ will have a similar dependence on $\beta$ (at least for small $\beta$), albeit possibly with somewhat different coefficients. 

From time series of the $T_i(t)$ taken at intervals of a few seconds and covering at least one day we determined the cross-correlation functions  $C^{i,j}(\tau)$ corresponding to signals at azimuthal positions displaced around the circle by $\pi$ ({\it i.e.} $j = i + 4$). These functions are given by 
\begin{equation}
C^{i,j}(\tau) = \langle [T_i(t) - \langle T_i(t)\rangle_t]\times[T_j(t+\tau) -  \langle T_j(t)\rangle_t] \rangle_t\ .
\label{eq:corfunc}
\end{equation}
We also calculated the auto-correlation functions corresponding to $i = j$ in Eq.~\ref{eq:corfunc}, for all eight thermometers.

Initially each sample was carefully leveled so that the tilt angle relative to gravity was less than $10^{-3}$ radian. Later it was tilted deliberately to study the influence of a non-zero $\beta$ on the heat transport.

The fluid was water at 40$^\circ$C where $\alpha = 3.88\times 10^{-4}$ K$^{-1}$,  $\kappa = 1.53\times 10^{-3}$ cm$^2$/s, and $\nu = 6.69\times 10^{-3}$  cm$^2$/s, yielding $\sigma = 4.38$.

\begin{figure}
\centerline{\psfig{file=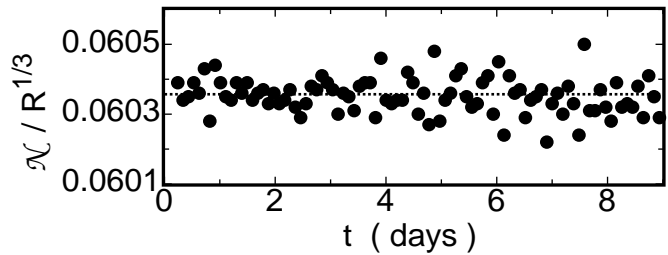,width=3.5in}}
\caption{The reduced Nusselt number ${\cal N}/R^{1/3}$ for $R = 9.43\times 10^{10}$ for the large sample as a function of time during a single experimental run that lasted nine days. Each data point is based on temperature and heat-current averages over a time interval of two hours. The dotted line corresponds to the value estimated by using all the data. Note that the entire vertical axis covers only a change of 0.8\%. The mean value is 0.06035, and the standard deviation from the mean is $5.1\times 10^{-5}$ or 0.084\%.}
\label{fig:N_of_t}
\end{figure}

\section{The Nusselt number of a vertical sample}
\label{sec:2}

\subsection{Initial transients}

In Fig.~\ref{fig:transients}a we show the initial evolution of the top and bottom temperatures of the large sample in a typical experiment. Initially the heat current was near zero and $T_b$ and $T_t$ were close to 40$^\circ$C. The sample had been equilibrated under these conditions for over one day. Near $t = 0.6$ h a new temperature set point of 50$^\circ$C was specified for the bottom plate, and the circulator for the top plate was set to provide $T_t \simeq 30^\circ$C.  From Fig.~\ref{fig:transients}a one sees that there were transients that lasted until about 0.9 h (1.2 h) for $T_b$ ($T_t$). These transients are determined by the response time and power capability of the bottom-plate heater and the top-plate cooling water and are unrelated to hydrodynamic phenomena in the liquid. Figures~\ref{fig:transients}b and c show the evolution of the heat current. After the initial rapid rise until $t \simeq 0.8$ h the current slowly evolved further to a statistically stationary value until $t \simeq 3$ h. A fit of the exponential function  $Q(t) = Q_\infty - \Delta Q ~ exp(-t/\tau_Q)$ to the data for $t > 1.2$ h is shown by the solid line in Fig.~\ref{fig:transients}c and yielded a relaxation time $\tau_Q = 0.48 \pm 0.04$ h. We attribute this transient to the evolution of the fluid flow. It is interesting to compare $\tau_Q$ with intrinsic time scales of the system. The vertical thermal diffusion time $\tau_v \equiv L^2/\kappa$ is 467 hours. Obviously it does not control the establishment of the stationary state. If we consider that it may be reduced by a factor of $1/{\cal N}$ with ${\cal N} = 263$, we still obtain a time sale of 1.78 hours that is longer than  $\tau_Q$. We believe that the relatively rapid equilibration is associated with the establishment of the top and bottom boundary layers that involve much shorter lengths $\it l_t$ and $\it l_b$. It also is necessary for the large-scale circulation to establish itself; but, as we shall see, its precise Reynolds number is unimportant for the heat transport.  In addition, the LSC can be created relatively fast since this is not a diffusive process. 

\begin{figure}
\centerline{\psfig{file=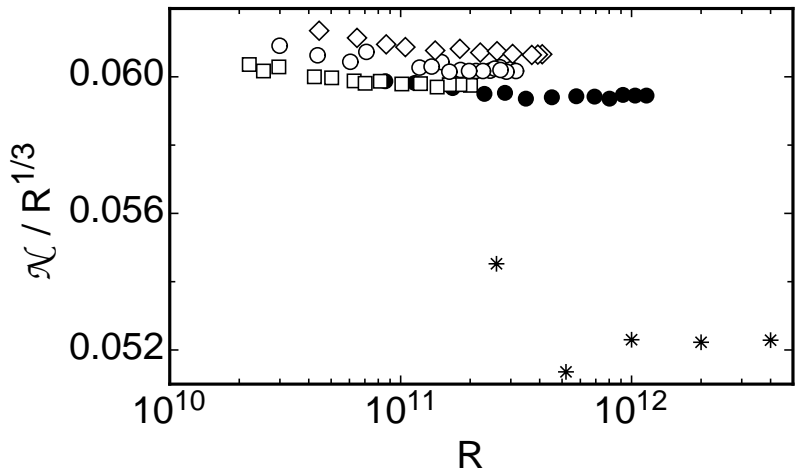,width=4.2in}}
\caption{The reduced Nusselt number ${\cal N}/R^{1/3}$ as a function of the Rayleigh number $R$. Stars: data from \cite{CRCC04} for $\sigma = 2.3$ and $\Gamma = 0.5$. Solid circles: Data from \cite{NBFA05} for $\sigma = 4.38$ and $\Gamma = 0.43$. Open symbols: form \cite{NBFA05} for $\Gamma = 0.67$.  Open squares: $\sigma = 5.42$. Open circles:  $\sigma = 4.38$. Open diamonds: $\sigma = 3.62$.}
\label{fig:N_compare}
\end{figure}

\subsection{Results under statistically stationary conditions}

Figure~\ref{fig:transients} shows the behavior of the system only during the first six hours and does not exclude the slow transients reported by Chill\`a {\it et al.} that occurred over time periods of ${\cal O}(10^2)$ hours. Thus we show in Fig.~\ref{fig:N_of_t} results for ${\cal  N}/R^{1/3}$ from a run using the large sample that was continued under constant externally imposed conditions for nine days.  Each point corresponds to a value of $\cal N$ based on a time average over two  hours of the plate temperatures and the heat current. Note that the vertical range of the entire graph is only 0.8 \%. Thus, within a small fraction of 1 \%, the results are time independent. Indeed, during nearly a year of  data acquisition for a $\Gamma = 1$ sample at various Rayleigh numbers, involving individual runs lasting from one to many days,  we have never experienced long-term drifts or changes of $\cal N$ after the first few hours. This differs dramatically from the observations of Chill\`a {\it et al}. who found changes by about 2 \% over about 4 days. We conclude that the slow transients observed by them for their $\Gamma = 0.5$ sample do not occur for $\Gamma \simeq 1$.

To document further the stationary nature of the system, we compared results from the large sample for $\cal N$ obtained from many runs, each of one to ten days' duration, over a period of about five months [\cite{NBFA05,FBNA05}].  The scatter of the data at a given $R$ is only about 0.1\%. This excellent reproducibility would not be expected if there were slow  transients due to transitions between different states of the LSC.

 Although work in our laboratory with other aspect ratios has been less extensive, we also have not seen any evidence of drifts or transients for the larger $\Gamma = 1.5, 2, 3,$ and 6 [\cite{FBNA05,BNFA05a}] nor for the smaller $\Gamma = 0.67, 0.43,$ and 0.28 [\cite{NBFA05,BNFA05a}]. It may be that $\Gamma = 0.5$, being near the borderline between a single-cell LSC and more complicated LSC structures [\cite{VC03,SV05,SXX05}], is unique in this respect.

In Fig.~\ref{fig:N_compare} we compare results for ${\cal N}/R^{1/3}$ from our large sample [\cite{NBFA05}] with those reported by Chill\`a {\it et al.} (stars). Our results are larger by about 15\%. To find a reason for this difference, we first look at the $\Gamma$ and $\sigma$ dependence. The open (solid) circles represent our data for $\sigma = 4.38$ and $\Gamma = 0.67~(0.43)$ and show that the dependence of $\cal N$ on $\Gamma$ is not very strong. The open squares (diamonds) are our results for $\Gamma = 0.67$ and $\sigma = 5.42~(3.62)$ and indicate that $\cal N$ actually {\it increases} slightly with $\sigma$. Thus 
the lower values of $\cal N$ (compared to ours) obtained by Chill\'a et al. for $\sigma = 2.3$ and $\Gamma = 0.5$ can not be explained in terms of the $\Gamma$ and $\sigma$ dependence of $\cal N$. Some of the difference can be attributed to non-Boussinesq effects that tend to reduce $\cal N$ [\cite{FBNA05}]. However, for the largest $\Delta T$ used by Chill\'a et al. (31$^\circ$C) we expect this effect to be somewhat less than 1 \% [\cite{FBNA05}]. Finally, the effect of the finite conductivity of the top and bottom plates comes to mind. This can reduce $\cal N$ by several \% when $\Delta T$ is large [\cite{CCC02,Ve04,BNFA05a}], but it is difficult to say precisely by how much. It seems unlikely that this effect can explain the entire difference, particularly at the smaller $R$ (and thus $\Delta T$) where it is relatively small.  

\begin{figure}
\centerline{\psfig{file=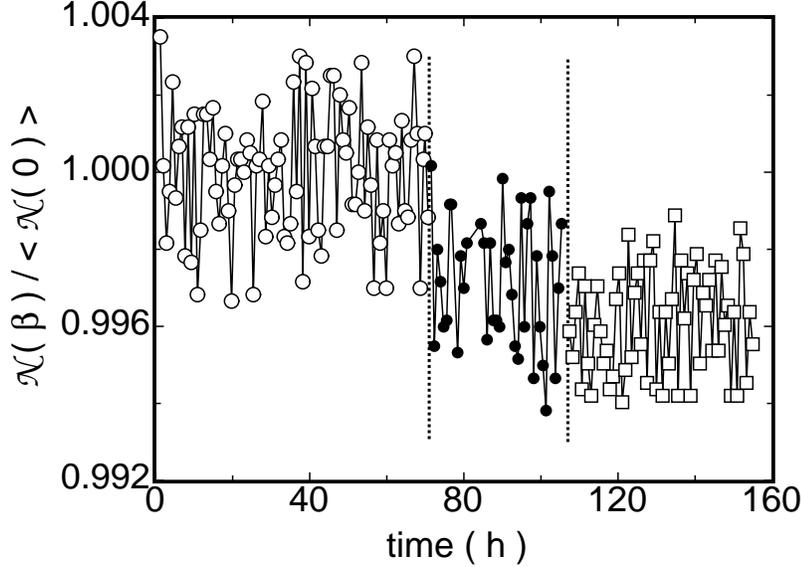,width=4.2in}}
\caption{The reduced Nusselt number ${\cal N}(\beta)/{\cal N}(0)$ as a function of time for the large sample and $R = 9.43\times 10^{10}$. Open circles: tilt angle $\beta = 0$. Solid circles: $\beta = 0.087$ radians. Open squares: $\beta = 0.122$ radians. All data are normalized by  the average  ${\cal N}(0)$ of the data for $\beta = 0$. Each data point is based on temperature and heat-current measurements over a two-hour period. The vertical dotted lines indicate the times when $\beta$ was changed.}
\label{fig:N_tiltresponse}
\end{figure}

\section{Tilt-angle dependence of the Nusselt number}

\begin{figure}
\centerline{\psfig{file=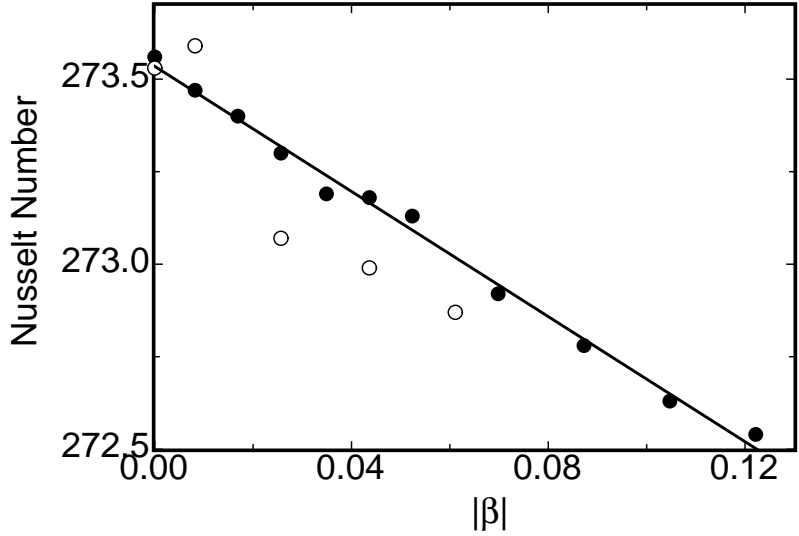,width=4.2in}}
\caption{The Nusselt number ${\cal N}(\beta)$ as a function of the tilt angle $\beta$ for the large sample with $R = 9.43\times 10^{10}$. Each point is the average over an entire run  of duration one day or longer. Solid circles:  $\beta > 0$. Open circles: $\beta < 0$. The solid line is a least-squares fit of a straight line (Eq.~\ref{eq:N}) to the data for $\beta > 0$.}
\label{fig:N_of_beta}
\end{figure}

\begin{figure}
\centerline{\psfig{file=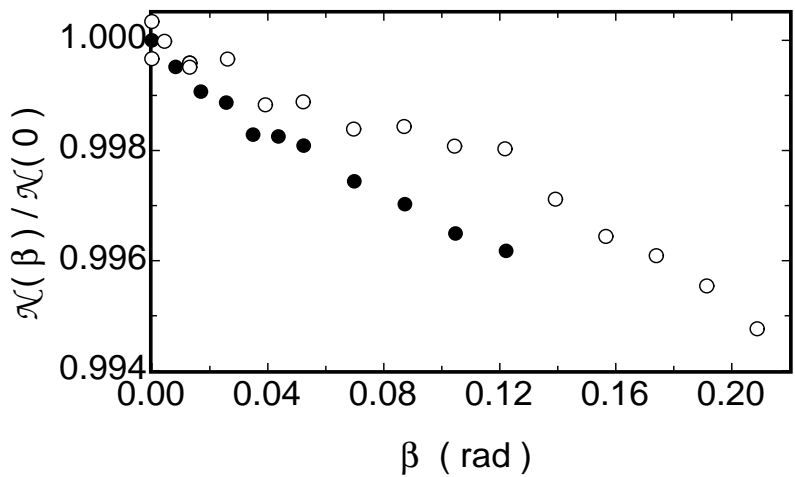,width=4.2in}}
\caption{The reduced Nusselt number ${\cal N}(\beta) /{\cal N}_0$ as a fucntion of the tilt angle $\beta$ for the large sample with $R = 9.43\times 10^{10}$ (solid circles) and the medium sample with $R = 1.13\times 10^{10}$ (open circles). }
\label{fig:Nred_of_beta}
\end{figure}

In Fig.~\ref{fig:N_tiltresponse} we show results for $\cal N$ from the large sample at $R = 9.43\times 10^{10}$. Each data point was obtained from a two-hour average of measurements of the various temperatures and of $Q$. Three data sets, taken in temporal succession,  for tilt angles $\beta = 0.000, ~0.087$, and 0.122 are shown. All data were normalized by the mean of the results for $\beta = 0$. Typically, the standard deviation from the mean of the data at a given $\beta$ was 0.13\%. The vertical dotted lines and the change in the data symbols show where $\beta$ was changed. One sees that tilting the cell caused a small but measurable reduction on $\cal N$. In Fig.~\ref{fig:N_of_beta} we show the mean value  for each tilt angle, obtained from runs of at least a day's duration a each $\beta$, as a function of $|\beta|$. One sees that  $\cal N$ decreases linearly with $\beta$. A fit of a straight line to the data yielded
\begin{equation}
{\cal N}(\beta) = {\cal N}_0  [1 - (3.1\pm 0.1)\times 10^{-2}|\beta|] \ .
\label{eq:N}
\end{equation}
with ${\cal N}_0 = 273.5$. Simlar results for the medium cell are compared with the large-cell results in Fig.~\ref{fig:Nred_of_beta}. At the smaller Rayleigh number of the medium sample the effect of $\beta$  on $\cal N$ is somewhat less. Because the effect of $\beta$ on $\cal N$  is so small, we did not make a more detailed investigation of its Rayleigh-number dependence.

Chill\`a et al. proposed a model that predicts a significant tilt-angle effect on $\cal N$ for $\Gamma = 0.5$ where they assume the existence of two LSC cells, one above the other. They also assumed that there would be no effect for $\Gamma = 1$ where there is only one LSC cell. Although we found an effect for our $\Gamma = 1$ sample, we note that it is a factor of about 50 smaller than the effect observed by Chill\`a et al. for $\Gamma = 0.5$.

\begin{figure}
\centerline{\psfig{file=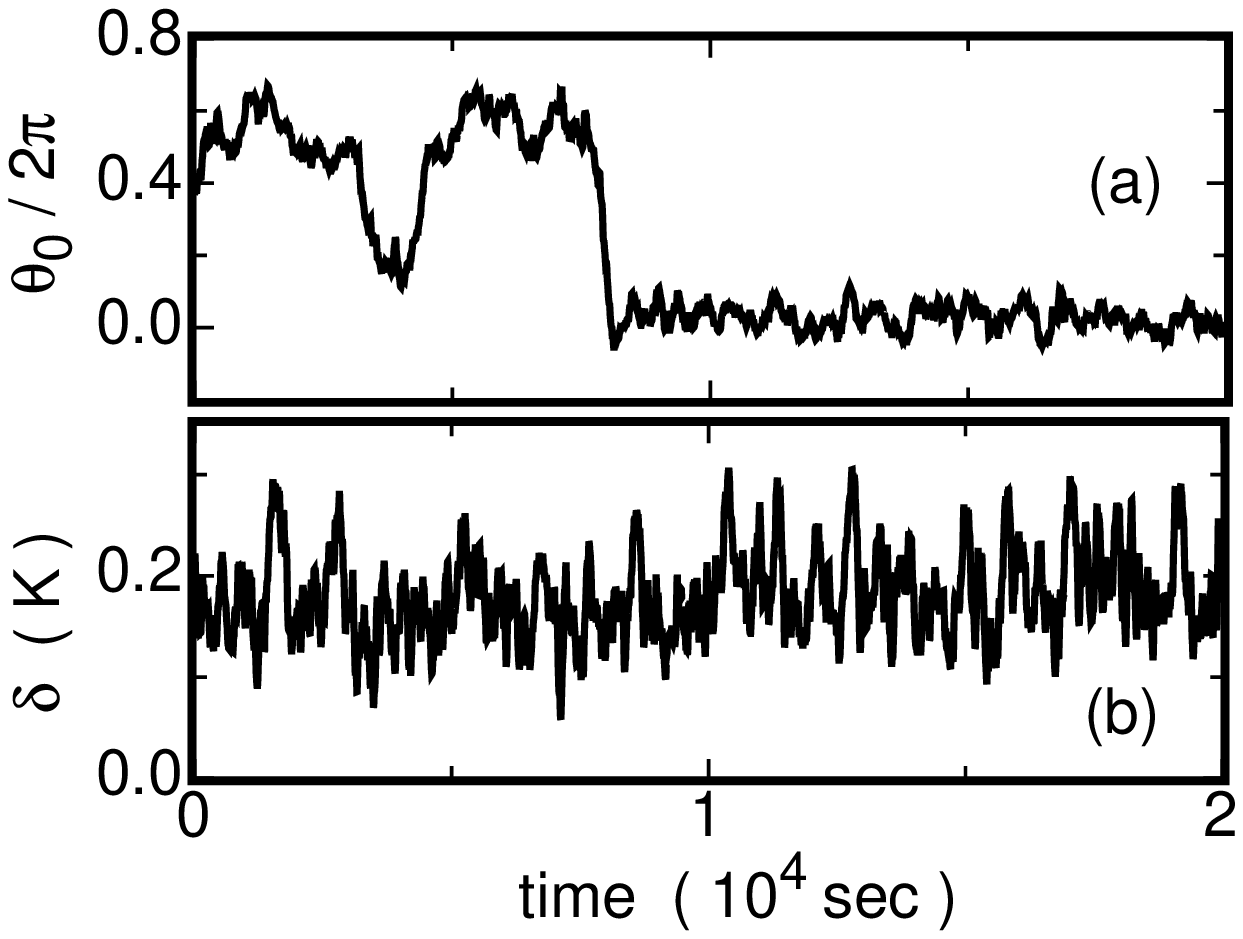,width=4.2in}}
\caption{(a): The orientation $\theta_0$ of the plane of circulation and (b): the temperature amplitude $\delta$ of the large-scale circulation as a function of time for the large sample and $R = 9.43\times 10^{10}$. At $t = 8000$ sec the sample was tilted by an angle $\beta = 0.087$ radians relative to gravity.}
\label{fig:theta+delta}
\end{figure}

\section{Tilt-angle dependence of the large-scale circulation}
\label{sec:3}

\subsection{The orientation}

\begin{figure}
\centerline{\psfig{file=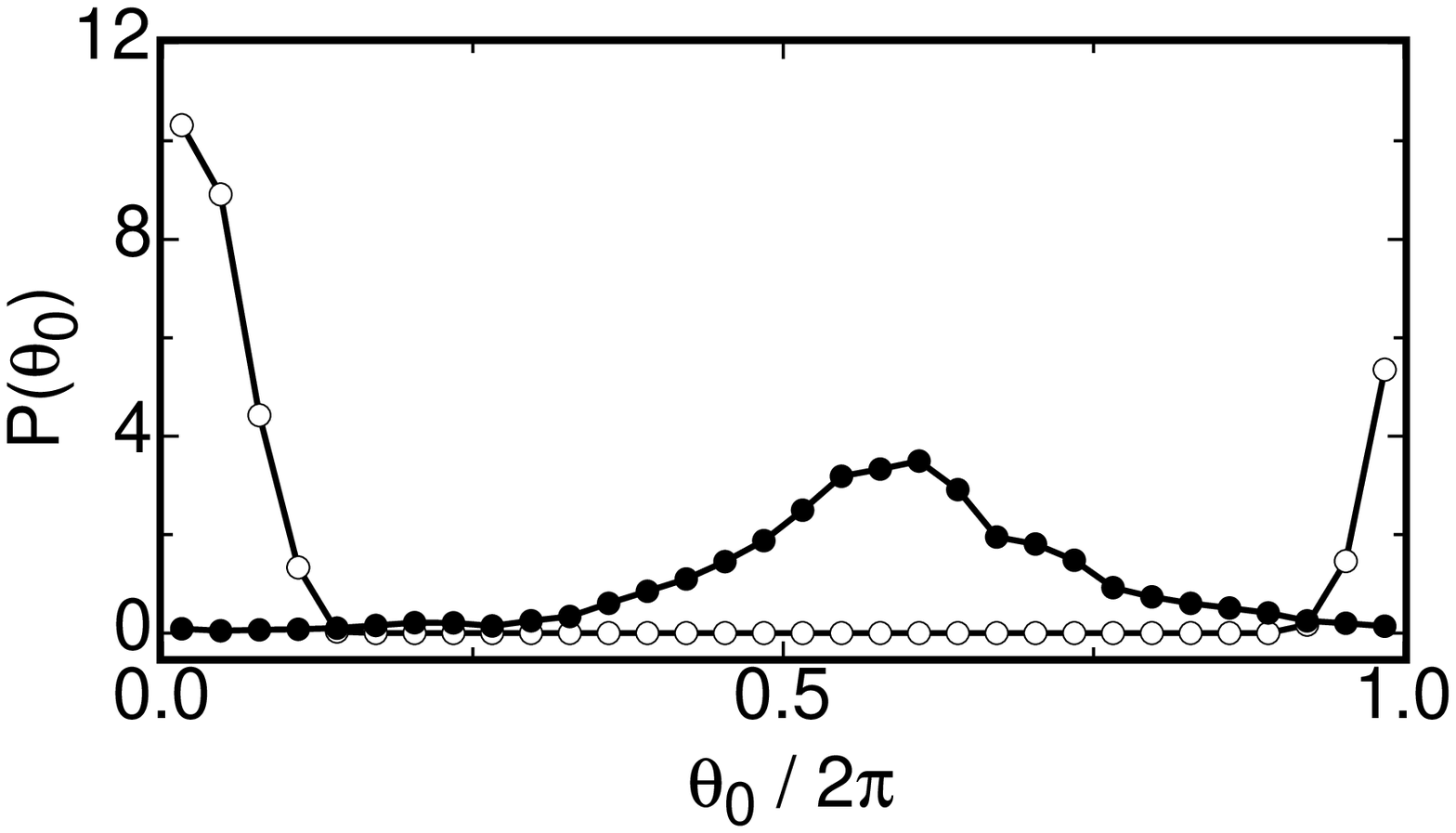,width=4.2in}}
\caption{The probability distribution $P(\theta_0)$ of the orientation $\theta_0$  of the plane of circulation of the large-scale flow for the large sample and $R = 9.43\times 10^{10}$ as a function of $\theta_0$. Solid circles: tilt angle $\beta = 0 \pm 0.001$. Open circles: $\beta = 0.087$ radians.}
\label{fig:P_of_theta}
\end{figure}

In Fig.~\ref{fig:theta+delta} we show the angular orientation $\theta_0$ (a) and the temperature amplitude $\delta$ (b) of the LSC. For the first 8000 sec shown in the figure, the sample was level ($\beta = 0.000\pm 0.001$). One sees that $\theta_0$ varied irregularly with time. The probability-distribution function $P(\theta_0)$ is shown in Fig.~\ref{fig:P_of_theta} as solid dots. Essentially all angles are sampled by the flow, but there is a preferred direction close to $\theta_0/2\pi = 0.6$. At t = 8000 sec, the sample was tilted through an angle $\beta = 0.087$ radian. The direction of the tilt was chosen deliberately so as to oppose the previously prevailing preferred orientation.  As a consequence one sees a sharp transition with a change of $\theta_0$ by approximately $\pi$. The temperature amplitude $\delta$ on average increased slightly, and certainly remained non-zero. From this we conclude that the transition took place via rotation of the LSC, and not by cessation that  would have involved a reduction of $\delta$ to zero [see \cite{BNA05b}]. We note that $\theta_0(t)$ fluctuated much less after the tilt. The results for $P(\theta_0)$ after the tilt are shown in Fig.~\ref{fig:P_of_theta} as open circles. They confirm  that the maximum was shifted close to $\theta_0 = 0$, and that the distribution was much more narrow.

\begin{figure}
\centerline{\psfig{file=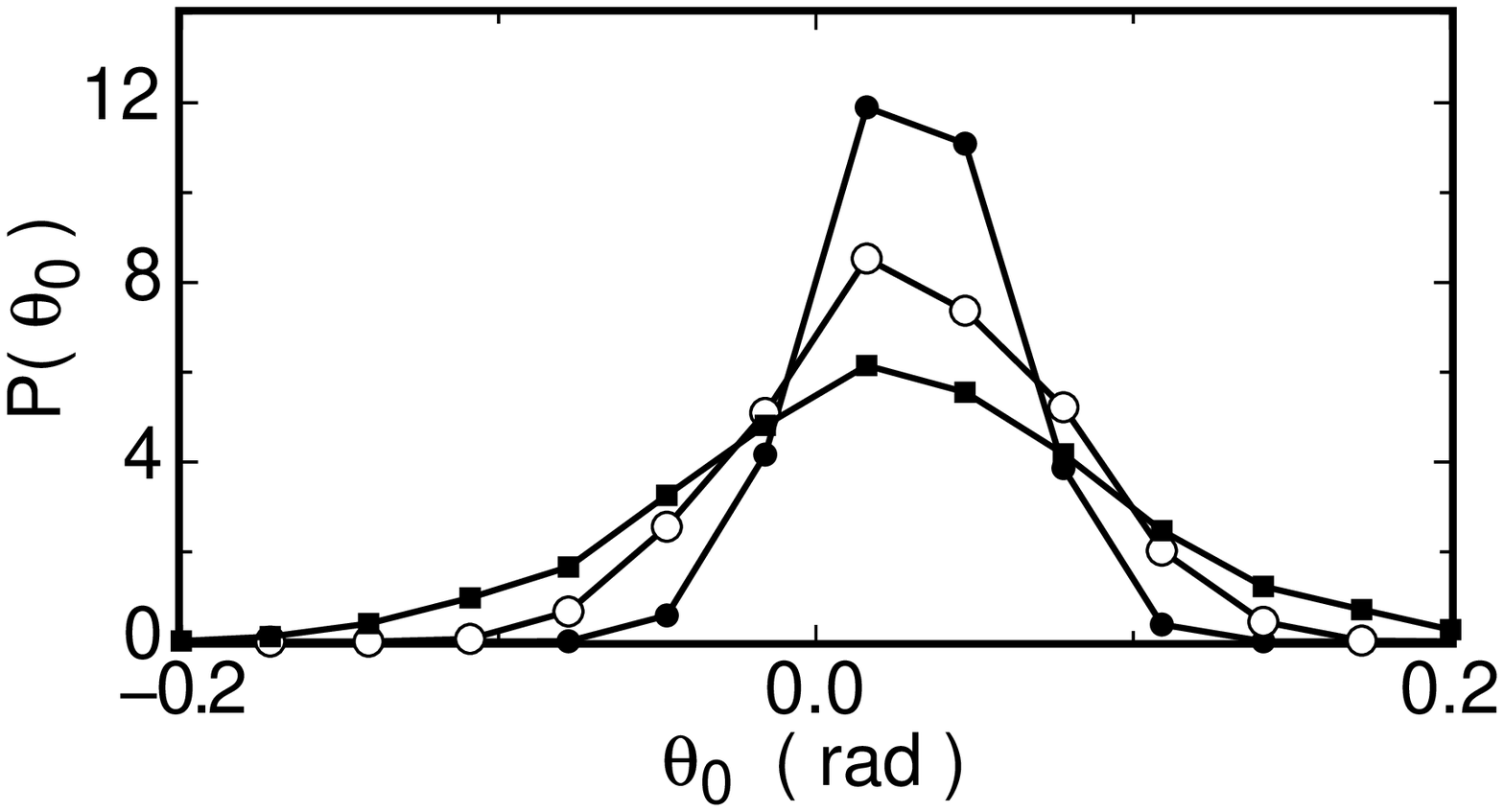,width=4.2in}}
\caption{The probability distribution $P(\theta_0)$ of the orientation $\theta_0$  of the plane of circulation of the large-scale flow for the large sample and $R = 9.43\times 10^{10}$ at three tilt angles $\beta$. Solid circles: $\beta = 0.122$. Open circles: $\beta = 0.044$. Solid squares: $\beta = 0.026$.}
\label{fig:Pshift_of_theta}
\end{figure}

\begin{figure}
\centerline{\psfig{file=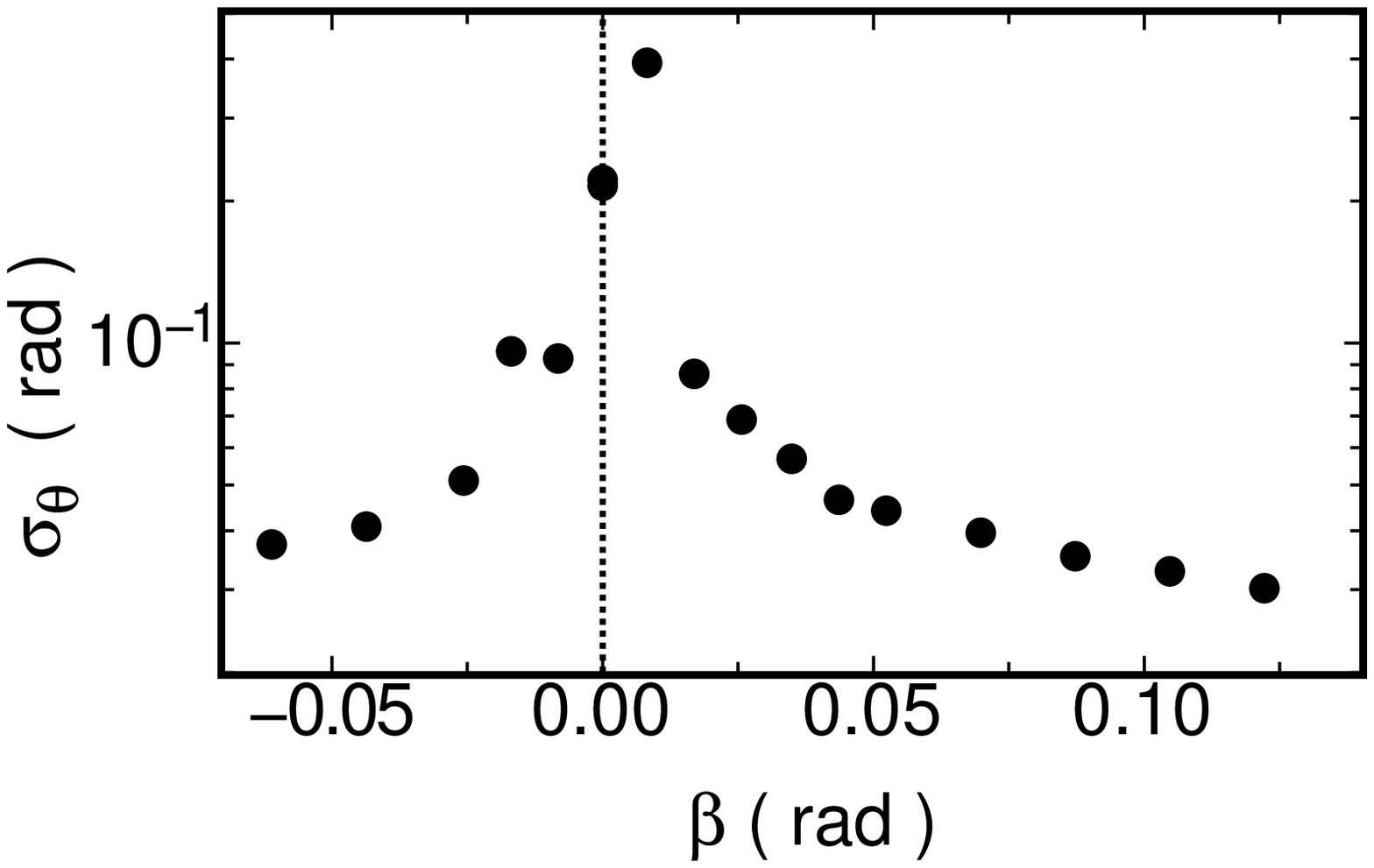,width=4.2in}}
\caption{The square root of the variance $\sigma_{\theta}$ of the probability distribution $P(\theta_0)$ of the orientation $\theta_0$  of the plane of circulation of the large-scale flow for the large sample and $R = 9.43\times 10^{10}$ as a function of the tilt angle $\beta$.}
\label{fig:dP_of_theta}
\end{figure}

In Fig.~\ref{fig:Pshift_of_theta} we show $P(\theta_0)$ for $\beta = 0.122,$ (solid circles),   0.044 (open circles), and 0.026 (solid squares). One sees that a reduction of $\beta$ leads to a broadening of $P(\theta_0)$. The square root of the variance of data like those in Fig.~\ref{fig:Pshift_of_theta} is shown in Fig.~\ref{fig:dP_of_theta} on a logarithmic scale as a function of $\beta$ on a linear scale. Even a rather small tilt angle caused severe narrowing of $P(\theta_0)$.

\begin{figure}
\centerline{\psfig{file=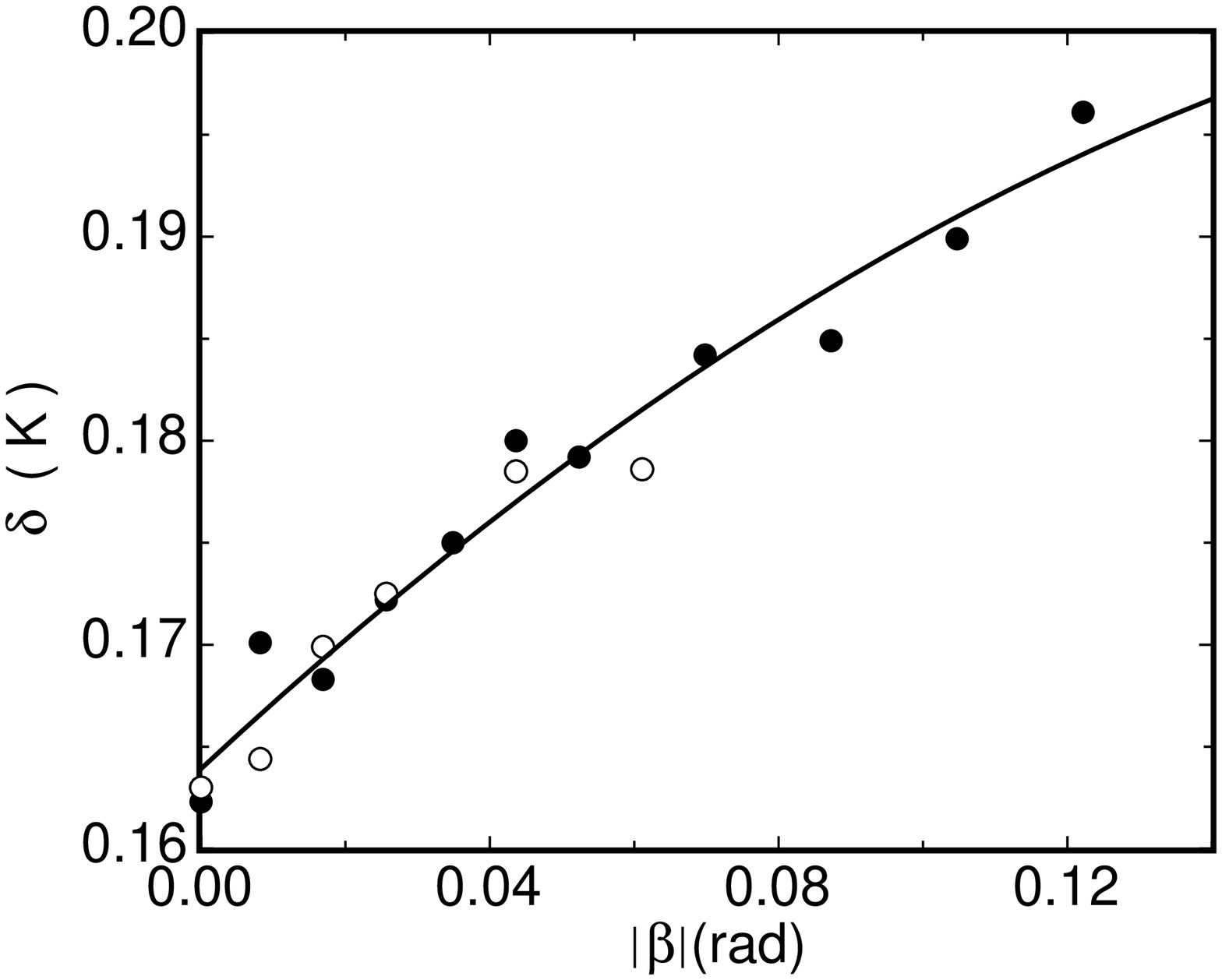,width=3.5in}}
\caption{The time-averaged temperature amplitude $\delta(\beta)$ of the LSC for the large sample  as a function of $|\beta|$. Solid circles: $\beta \geq 0$. Open circles: $\beta < 0$.  For this example $R = 9.43\times 10^{10}$.}
\label{fig:delta}
\end{figure}

\subsection{The temperature amplitude}

In Fig.~\ref{fig:delta} we show the temperature-amplitude $\delta(\beta)$ of the LSC as a function of $\beta$. As was the case for $\cal N$, the data are averages over the duration of a run at a given $\beta$ (typically a day or two). The solid (open) circles are for positive (negative) $\beta$. The data can be represented well by either a linear or a quadratic equation. A least-squares fit yielded
\begin{equation}
\delta(\beta) = \delta(0)\times [1 + (1.84 \pm 0.45)  |\beta| - (3.1 \pm 3.9) \beta^2]
\label{eq:delta}
\end{equation}
with $ \delta(0) = 0.164$ K.

\subsection{The Reynolds numbers}

\begin{figure}
\centerline{\psfig{file=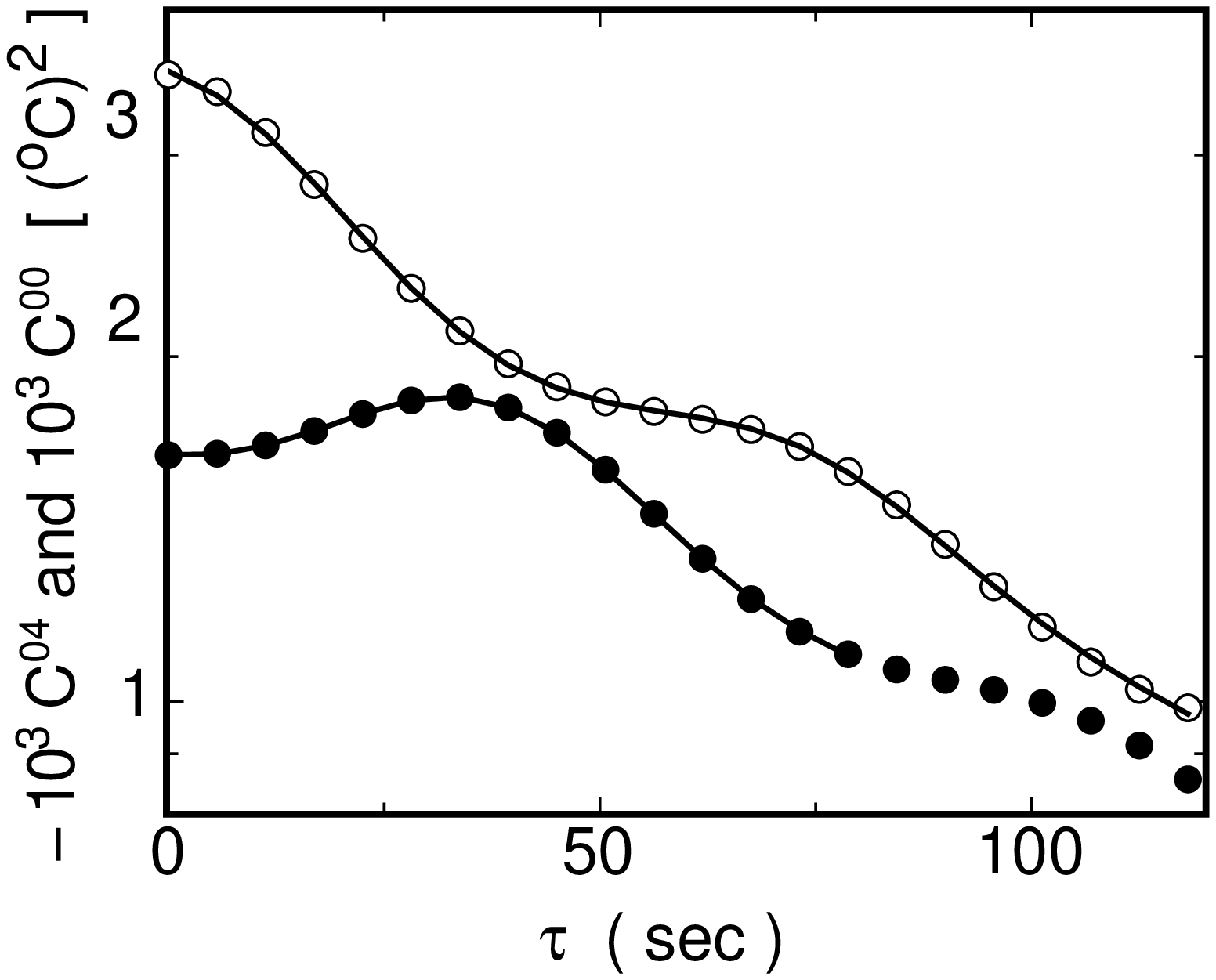,width=4.2in}}
\caption{The cross-correlation function $C^{04}(\tau)$ between thermometers 0 and 4 and the auto-correlation function  $C^{00}(\tau)$ of thermometer 0 for the large sample. The solid lines are fits of Eqs.~\ref{eq:ccfit} and \ref{eq:acfit} to the data. They also indicate the range of $\tau$ used for the fits. For this example the tilt angle was $\beta = -0.009$ and $R = 9.43\times 10^{10}$.}
\label{fig:cor}
\end{figure}

\begin{figure}
\centerline{\psfig{file=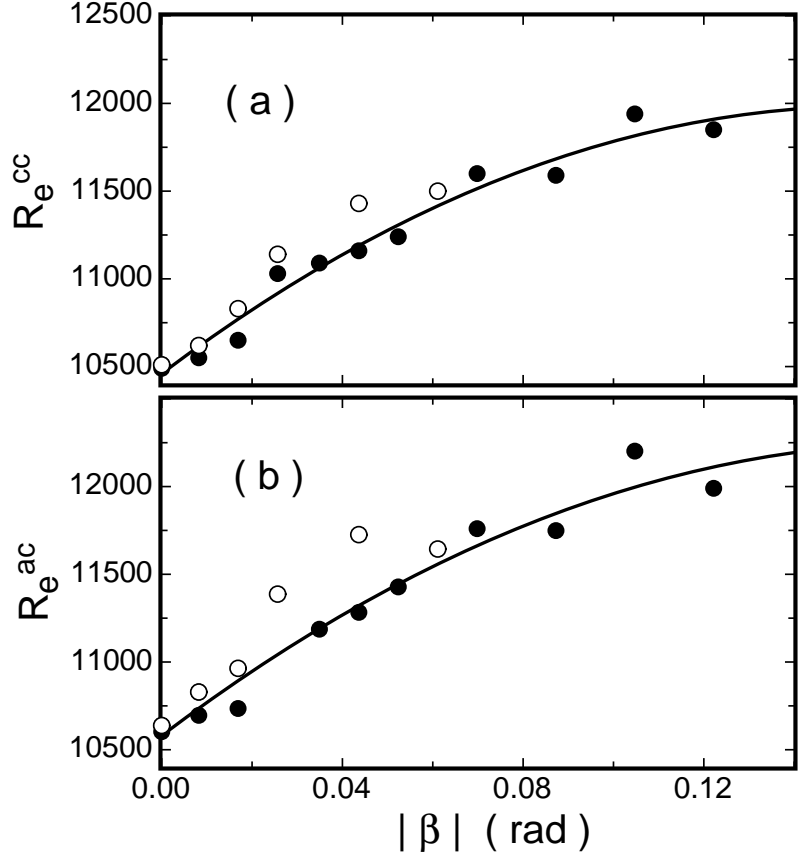,width=4.2in}}
\caption{(a): The Reynolds number $R_e^{cc}(|\beta|)$ obtained from the temperature cross-correlation functions, and (b): the Reynolds number $R_e^{ac}(|\beta|)$ obtained from the temperature auto-correlation functions,  of the LSC for the large sample and $R = 9.43\times 10^{10}$ as a function of the absolute value $|\beta|$ of the tilt angle. Solid circles: $\beta \geq 0$. Open circles: $\beta < 0$. For this example $R = 9.43\times 10^{10}$.}
\label{fig:Re}
\end{figure}

\begin{figure}
\centerline{\psfig{file=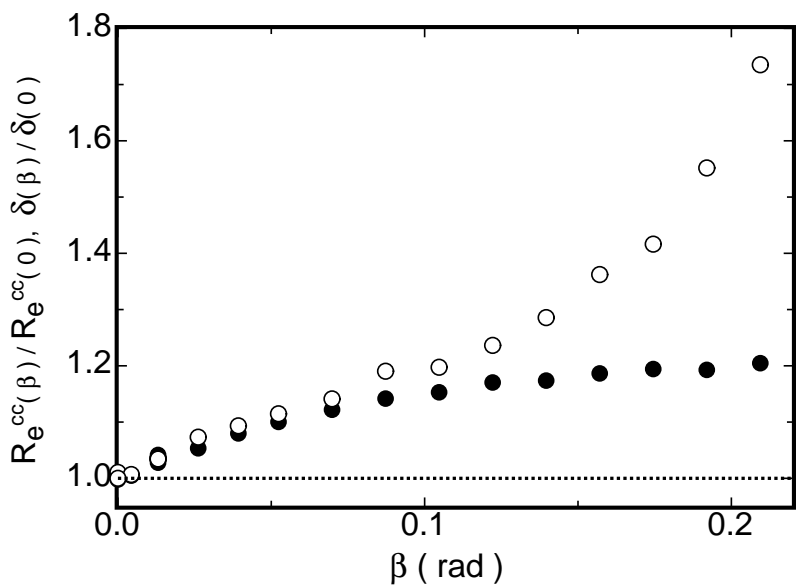,width=4.2in}}
\caption{The Reynolds-number ratio $R_e^{cc}(\beta)/R_e^{cc}(0)$ (solid circles) obtained from the temperature cross-correlation functions, and the amplitude ratio $\delta(\beta)/\delta(0)$ (open circles),  of the LSC of the medium sample for $R = 1.13\times 10^{10}$ as a function of $\beta$.}
\label{fig:Re_delta_ratios}
\end{figure}

\begin{figure}
\centerline{\psfig{file=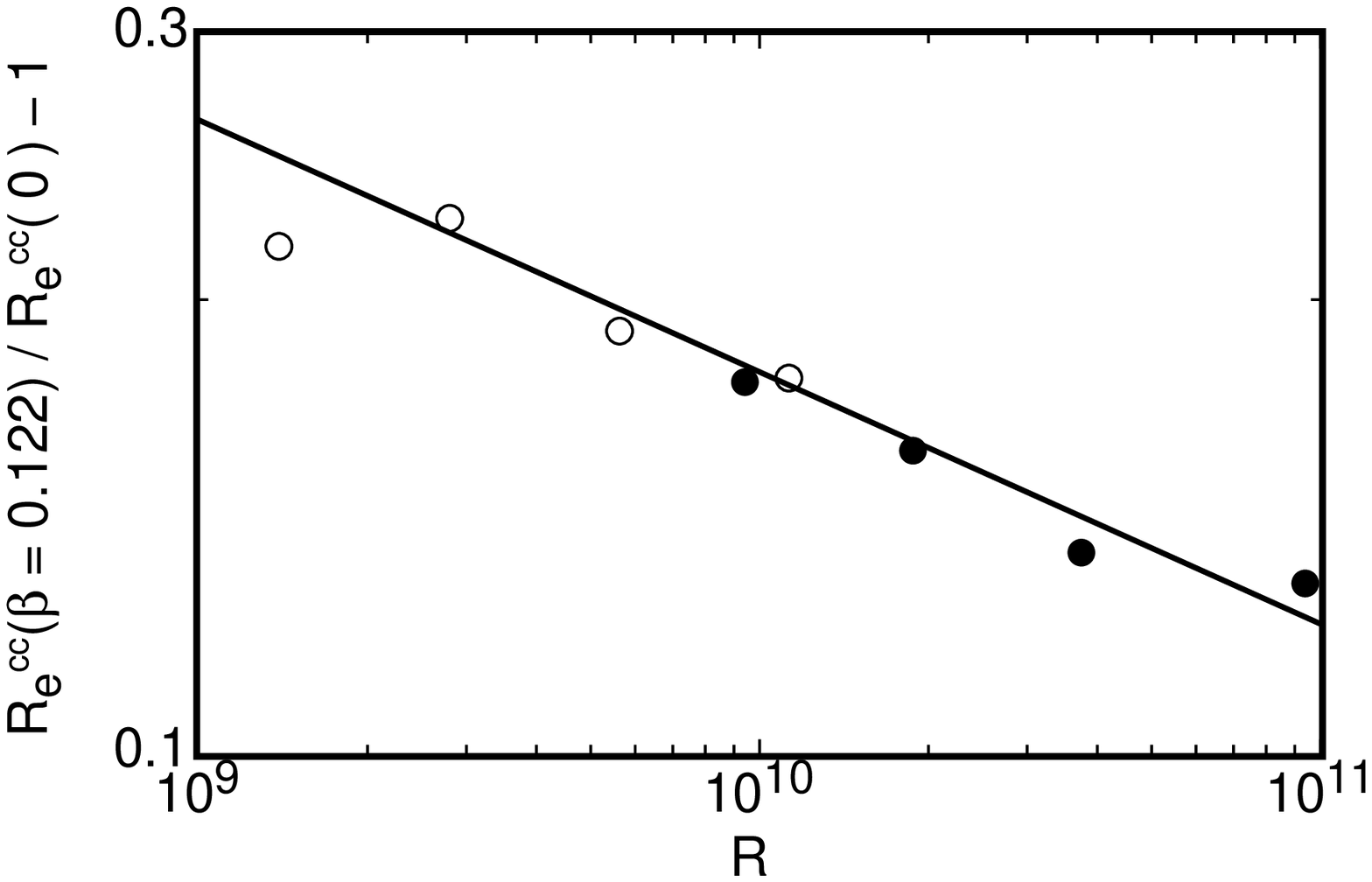,width=4.2in}}
\caption{The Reynolds-number ratio $R_e^{cc}(\beta = 0.122)/R_e^{cc}(0)$ of the LSC as a function of $R$. Open circles: medium sample. Solid circles: large sample. Solid line: powerlaw with an exponent of $1/6$.}
\label{fig:Re_ratios}
\end{figure}

Using Eq.~\ref{eq:corfunc}, we calculated the auto-correlation functions  (AC) $C^{i,i}, i = 0, \ldots, 7$, as well as the  cross-correlation functions (CC) $C^{i,j}, j = (i + 4) \% 8, i = 0, \ldots, 7$, of the temperatures measured on opposite sides of the sample. Typical examples are shown in Fig.~\ref{fig:cor}. The CC  has a characteristic peak that we associate with the passage of relatively hot or cold volumes of fluid at the thermometer locations. Such temperature cross-correlations have been shown, {\it e.g.} by \cite{QT01b} and \cite{QT02},  to yield delay times equal to those of velocity-correlation measurements, indicating that warm or cold fluid volumes travel with the LSC. 
The function  
\begin{equation}
C^{i,j}(\tau) = - b_0 exp \left ( -\frac{\tau}{\tau^{i,j}_0} \right ) - b_1 exp \left [ -\left ( \frac{\tau - t^{i,j}_1}{\tau^{i,j}_1}\right )^2 \right]\ ,
\label{eq:ccfit}
\end{equation}
consisting of an exponentially decaying background (that we associate with the random time evolution of $\theta_0$) and a Gaussian peak, was fitted to the data for the CC. The fitted function is shown in Fig.~\ref{fig:cor} as a solid line over the range of $\tau$ used in the fit. It is an excellent representation of the data and yields the half turnover time ${\cal T}/2 = t^{i,j}_1$ of the LSC. 
Similarly, we fitted the function
\begin{equation}
C^{i,i}(\tau) = b_0 exp\left (- \frac{\tau}{\tau^{i,i}_0}\right ) + b_1 exp \left [- \left (\frac{\tau}{\tau^{i,i}_1} \right )^2 \right ] + b_2 exp \left [-\left ( \frac{\tau - t^{i,i}_2}{\tau^{i,i}_2}\right )^2 \right ]
\label{eq:acfit}
\end{equation}
to the AC data. It consists of two Gaussian peaks, one centered at $\tau = 0$ and the other at $\tau = t^{i,i}_2$, and the exponential background. We interpret the location $ t^{i,i}_2$ of the second Gaussian peak to correspond to a complete turnover time $\cal T$ of the LSC.

In terms of the averages $<{t^{i,j}_1}>$ and $<{t^{i,i}_2}>$ over all 8 thermometers or thermometer-pair combinations we define [\cite{QT02,GL02}] the Reynolds numbers
\begin{equation}
R^{cc}_{e} =  (L/ <t^{i,j}_1>)  (L/\nu)
\label{eq:Re_cc}
\end{equation}
and
\begin{equation}
R^{ac}_{e} =  (2L/ <t^{i,i}_2>)  (L/\nu)\ .
\label{eq:Re_ac}
\end{equation}
Here the length scale $2L$ was used to convert the turnover time ${\cal T}$ into a LSC speed $2L/{\cal T}$. For $\Gamma = 1$, the length $4L$ might have been used instead, as was done for instance by \cite{LSZX02}. This would have led to a Reynolds number larger by a factor of two. In Fig.~\ref{fig:Re}a and b we show $R_e^{cc}(|\beta|)$ and  $R_e^{ac}(|\beta|)$ respectively. The solid circles are for positive and the open ones for negative $\beta$. One sees that $R_e^{cc}(|\beta|)$ and  $R_e^{ac}(|\beta|)$ initially grow linearly with $\beta$, but the data also reveal some curvature as $|\beta|$ becomes larger. Thus we fitted quadratic equations  to the data and obtained
\begin{equation}
R_e^{cc}(\beta) = R_e^{cc}(0)\times [1 + (1.85\pm 0.21)  |\beta| - (5.9\pm 1.7)  \beta^2]
\label{eq:Recc}
\end{equation}
and
\begin{equation}
R_e^{ac}(\beta) = R_e^{ac}(0)\times [1 + (1.72\pm 0.38)  |\beta| - (4.1\pm 3.2)  \beta^2]
\label{eq:Reac}
\end{equation}
with $R_e^{cc}(0) = 10467 \pm 43$ and $R_e^{ac}(0) = 10565 \pm 82$ (all parameter errors are 67\% confidence limits). The results for $R_e^{cc}(0)$ and $R_e^{ac}(0)$ are about 10\% higher than the prediction by \cite{GL02} for our $\sigma$ and $R$. The excellent agreement between $R_e^{cc}$ and  $R_e^{ac}$ is consistent with the idea that the CC yields ${\cal T}/2$ and that the AC gives ${\cal T}$.
As expected (see Sect.~\ref{sec:apparatus}), the $\beta$-dependences of both Reynolds numbers are the same within their uncertainties. It is interesting to see that the coefficients of the linear term also agree with the corresponding coefficient for $\delta$ (Eq.~\ref{eq:delta}). This suggests that there may be a closer relationship between $\delta $ and $R_e$ than we would have expected {\it a priori}.
However, the coefficient of the linear term in Eq.~\ref{eq:Recc} or \ref{eq:Reac} is larger by a factor of about 50 than the corresponding coefficient for the Nusselt number in Eq.~\ref{eq:N}.

In Fig.~\ref{fig:Re_delta_ratios} we show measurements of $R_e^{cc}$ and of $\delta$, each normalized by its value at $\beta = 0$, as a function of $\beta$ for the medium
  sample and $R = 1.13\times 10^{10}$. For this sample we were able to attain larger values of $\beta$ than for the large one. One sees that $\delta$ and $R_e^{cc}$ have
 about the same $\beta$ dependence for small $\beta$,  but that $\delta$ then increases more rapidly than $R_e^{cc}$ as $\beta$ becomes large. Although we do not know the reason for this behavior, it suggests that the larger speed of the LSC enhances the thermal contact between the side wall and the fluid interior.

The Rayleigh-number dependence of $R_e^{cc}$ at constant $\beta$ is shown in Fig.~\ref{fig:Re_ratios}. Here the open (solid) circles are from the  medium (small) sample. There is consistency between the two samples, and the data can be described by a power law with a small negative exponent. The solid line is drawn to correspond to an exponent of $-1/6$. 

\subsection{A model for the enhancement of the Reynolds number}

As seen in Fig.~\ref{fig:P_of_theta}, the LSC assumes an orientation for which gravity enhances the velocity above (below) the bottom (top plate), {\it i.e.} the LSC flows ``uphill" at the bottom where it is relatively warm and ``downhill" at the top where it is relatively cold. This leads to an enhancement of the Reynolds number of the LSC.  As suggested by \cite{CRCC04}, one can model this effect by considering the buoyancy force per unit area parallel to the plates. This force can be estimated to be $\rho {\it l}g \beta \alpha \Delta T/2$ where ${\it l}$ is the boundary-layer thickness. It is opposed by the increase of the viscous shear stress across the boundary layer that may be represented by $\rho \nu u'/{\it l}$ where $u'$ is the extra speed gained by the LSC due to the tilt. Equating the two, substituting 
\begin{equation}
{\it l} = L / (2{\cal N})\ ,
\label{eq:BL}
\end{equation}
solving for $u'$, using Eq.~\ref{eq:R} for $R$, and defining $R_e' \equiv (L/\nu)u'$ one obtains 
\begin{equation}
R_e'=  \frac{R \beta}{8 \sigma {\cal N}^2}
\end{equation}
for the enhancement of the Reynolds number of the LSC. From our measurements at large $R$ we found that $R_e$ [\cite{ABFN05}] and $\cal N$ [\cite{NBFA05}] can be represented within experimental uncertainty by 
\begin{eqnarray}
R_e &=& 0.0345 R^{1/2}\ , 		\label{eq:Re_exp} \\
{\cal N} &=& 0.0602 R^{1/3}\ , 		\label{eq:Nu_exp}
\end{eqnarray}
giving
\begin{equation}
\frac{R_e'}{R_e} = 1.00\times 10^3 R^{-1/6} \sigma^{-1} \beta\ .
\label{eq:Re'}
\end{equation}
For our $\sigma = 4.38$ and $R = 9.43\times 10^{10}$ one finds $R_e'/R_e = 3.4 \beta$, compared to the experimental value $(1.9\pm 0.2) \beta$ from $R_e^{cc}$ [Eq.~\ref{eq:Recc}] and $(1.7\pm 0.4) \beta$ from $R_e^{ac}$ [Eq.~\ref{eq:Reac}]. We note that the coefficient $1.00\times 10^3$ in Eq.~\ref{eq:Re'} depends on the definition of $R_e$ given in Eqs.~\ref{eq:Re_cc} an \ref{eq:Re_ac} that was used in deriving the result Eq.~\ref{eq:Re_exp}. If the length scale $4L$ had been used instead of $2L$ to define the speed of the LSC, as was done for instance by \cite{LSZX02}, this coefficient would have been smaller by a factor of two, yielding near-perfect agreement with the measurements.  In Fig.~\ref{fig:Re_ratios} one sees that the predicted dependence on $R^{-1/6}$ also is in excellent agreement with the experimental results.  
However, such good agreement may be somewhat fortuitous, considering the approximations that were made in the model. Particularly the use of Eq.~\ref{eq:BL} for the boundary-layer thickness is called into question at a quantitative level by measurements of \cite{LX98} that revealed a significant variation of $\it l$ with lateral position. In addition, it is not obvious that the thermal boundary-layer thickness $\it l$ should be used, as suggested by \cite{CRCC04}, to estimate the shear stress; perhaps the thickness of the viscous BL would be more appropriate.

In discussing their $\Gamma = 0.5$ sample, \cite{CRCC04} took the additional step of assuming that the relative change due to a finite $\beta$ of $\cal N$ is equal to the relative change of $R_e$. For our sample with $\Gamma = 1$ this assumption does not hold.  As we saw above, the relative change of $\cal N$ is a factor of about 50 less than the relative change of $R_e$. The origin of the (small) depression of the Nusselt number is not so obvious. Naively one might replace $g$ in the definition of the Rayleigh number by $g~cos(\beta)$; but this would lead to a correction of order $\beta^2$ whereas the experiment shows that the correction is of order $\beta$, albeit with a coefficient that is smaller than of order one. The linear dependence suggests that the effect of $\beta$ on $\cal N$ may be provoked by the change of $R_e$ with $\beta$, but not in a direct causal relationship.

\begin{figure}
\centerline{\psfig{file=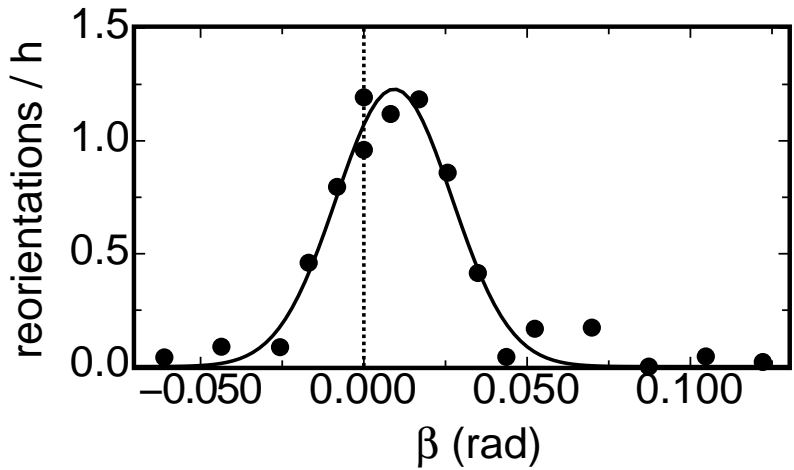,width=4.2in}}
\caption{The number of reorientation events per hour of the angular orientation of the plane of circulation of the LSC in the large sample for $R = 9.43\times 10^{10}$.}
\label{fig:n}
\end{figure}

\section{Tilt-angle dependence of reorientations of the large-scale circulation}

It is known from direct numerical simulation [\cite{HYK91}] and from several experiments [\cite{CCS97,NSSD01,SBN02,BNA05b}] that the LSC can undergo relatively sudden reorientations.  Not unexpectedly, we find that the tilt angle strongly influences the frequency of such events. For a level sample ($\beta = 0$) we demonstrated elsewhere [\cite{BNA05b}] that reorientations can  involve changes of the orientation of the plane of circulation of the LSC through any angular increment $\Delta \theta$, with the probability $P(\Delta \theta)$ increasing with decreasing $\Delta \theta$. Thus, in order to define a ``reorientation", we established certain criteria. We required that the magnitude of the net angular change $|\Delta\theta|$ had to be greater than $\Delta\theta_{min}= (2\pi)/8$. In addition we specified that the magnitude of the net average azimuthal rotation rate $|\dot\theta| \equiv |\Delta \theta / \Delta t|$ had to be greater than $\dot\theta_{min} = 0.1/{\cal T}$ where $\cal T$ is the LSC turnover time and $\Delta t$ is the duration of the reorientation (we refer to \cite{BNA05b} for further details). Using these criteria, we found that the number of reorientation events $n(\beta)$ at constant $R = 9.43\times 10^{10}$ decreased rapidly with increasing $|\beta|$. These results are shown in Fig.~\ref{fig:n}. It is worth noting that nearly all of these events are rotations of the LSC and very few involved a cessation of the circulation. A least-squares fit of the Gaussian function
\begin{equation}
n(\beta) = N_0 exp[-(\beta - \beta_0)^2/w^2]
\label{eq:N_r}
\end{equation}
to the data yielded $N_0= 1.23 \pm 0.06$ events per hour, $\beta_0 = 0.0093 \pm 0.0010$ rad, and $w = 0.0251 \pm 0.0015$ rad. It is shown by the solid line in the figure.

We note that the distribution function is not centered on $\beta = 0$. The displacement of the center by about 9 mrad is much more than the probable error of $\beta$. We believe that it is caused by the effect of the Coriolis force on the LSC that will be discussed in more detail elsewhere [\cite{BNA05c}].

\section{Tilt-angle dependence of the center temperature}

\begin{figure}
\centerline{\psfig{file=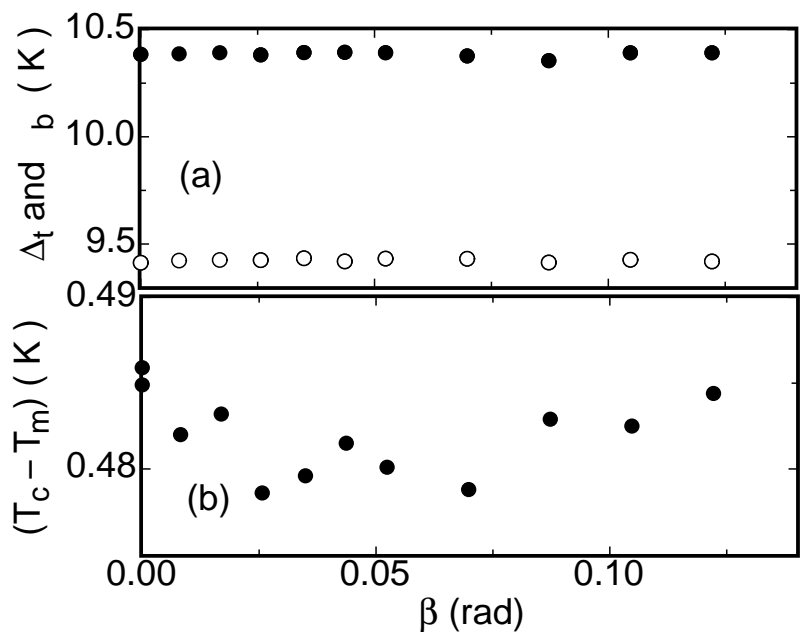,width=4.2in}}
\caption{(a): The temperature difference in Kelvin between the bottom and the center  ($\Delta_b = T_b - T_c$, solid circles) and the center and the top ($\Delta_t = T_c - T_t$, open circles) as a function of the tilt angle $\beta$. (b): The temperature difference $T_c - T_m = (\Delta_t - \Delta_b)/2$ between the center temperature $T_c$ and the mean temperature $T_m = (T_t + T_b)/2$. The center temperature is the average of the values given by the eight side-wall thermometers.}
\label{fig:NOB}
\end{figure}

We saw from Fig.~\ref{fig:delta} that the increase of $R_e$ with $\beta$ led to an increase of the amplitude $\delta$ of the azimuthal temperature variation at the horizontal mid-plane. An additional question is whether the tilt-angle effect on this system has an asymmetry between the top and bottom that would lead to a change of the mean center temperature $T_c$ (see Eq.~\ref{eq:T_i}). \cite{CRCC04} report such an effect for their $\Gamma = 0.5$ sample. For a Boussinesq sample with $\beta = 0$ we expect that $T_c = T_m$ with  $T_m = (T_t + T_b)/2$ ($T_t$ and $T_b$ are the top and bottom temperatures respectively), or equivalently that $\Delta _t = T_c - T_t$ is equal to $\Delta_b = T_b - T_c$.
 A difference between $\Delta_b$ and $\Delta_t$ will occur when the fluid properties have a significant  temperature dependence [\cite{WL91,ZCL97}], {\it i.e.} when there are significant deviations from the Boussinesq approximation.  
For the sequence of measurements with the large apparatus and $R = 9.43\times 10^{10}$ as a function of $\beta$ the mean value of $\Delta T = T_b - T_t$ was $19.808 \pm 0.018 {^\circ}$C and $T_c - T_m$ was 0.97$^\circ$C, indicating a significant non-Boussinesq effect. In Fig.~\ref{fig:NOB}a we show $\Delta_t$ and $\Delta_b$ as a function of $\beta$. One sees that increasing $\beta$ does not have a significant effect for our $\Gamma = 1$ sample. This is shown with greater resolution in Fig.~\ref{fig:NOB}b where $T_c - T_m = (\Delta_t - \Delta_b)/2$ is shown. We believe that the small variation, over a range of about $5\times 10^{-3} {^\circ}$C, is within possible systematic experimental errors and consistent with the absence of a tilt-angle effect.

\section{Summary}

In this paper we reported on an experimental investigation of the influence on turbulent convection of a small tilt angle $\beta$  relative to gravity of the axes of two cylindrical Rayleigh-B\'enard samples. The aspect ratios were $\Gamma \simeq 1$.

Where there was overlap, there were significant differences between our results and those obtained by \cite{CRCC04} for a $\Gamma = 0.5$ sample. We found our system to establish a statistically stationary state quickly, within a couple of hours, after a Rayleigh-number  change whereas \cite{CRCC04} found long transients that they attributed to changes of the LSC structure. We found a very small depression of the Nusselt number $\cal N$ with increasing $\beta$, by about 4\% per radian at small $\beta$. \cite{CRCC04} found a decrease by 200\% per radian for their sample. 

In contrast to the very  small effect of $\beta$ on $\cal N$, we found an increase of the Reynolds number $R_e$ by about 180\% per radian for small $\beta$. The small effect on $\cal N$ in the presence of this large change of $R_e$ indicates that the heat transport does not depend strongly on the speed of the LSC sweeping over the boundary layers. Instead, $\cal N$ must be determined by instability mechanisms of the boundary layers,  and the associated efficiency of the ejection of hot (cold) volumes (so-called ``plumes") of fluid from the bottom (top) boundary layer. 

It is interesting to note that the strong dependence of $R_e$ on $\beta$ in the presence of only a very weak dependence of $\cal N$ on $\beta$ can be accommodated quite well within the model of \cite{GL02}. The Reynolds number can be changed by introducing a $\beta$-dependence of the parameter $a(\beta)$ in their Eqs.~(4) and (6). As pointed out by them, a change of $a$ has no influence on the predicted value for $\cal N$. 

We also measured the frequency of rapid LSC reorientations that are known to occur for $\beta = 0$. We found  that such events are strongly suppressed by a finite $\beta$. Even a mild breaking or the rotational invariance, corresponding to $\beta \simeq 0.04$, suppresses re-orientations almost completely.

\section{Acknowledgment}

We are grateful to Siegfried Grossmann and Detlef Lohse for fruitful exchanges. This work was supported by the United States Department of Energy through Grant DE-FG02-03ER46080.


\begin{thebibliography}{}

\bibitem[Ahlers, Grossmann \& Lohse (2002)]{AGL02}{\sc Ahlers, G., Grossmann, S. \& Lohse, D.} 2002 Hochpr\"azision im Kochtopf: Neues zur turbulenten Konvektion. {\em Physik Journal} {\bf 1 (2)}, 31--37.

\bibitem[Ahlers {\em et~al.\/} (2005)]{ABFN05}{\sc Ahlers, G., Brown, E., Funfschilling, D., \& Nikolaenko, A.} 2005, unpublished.


\bibitem[Belmonte {\em et~al.\/} (1995)]{BTL95}{\sc Belmonte, A., Tilgner, A., \& Libchaber, A.} 1995 Turbulence and internal waves in side-heated convection. {\em Phys. Rev. E} {\bf 51}, 5681 -- 5687.


\bibitem[Brown {\em et~al.\/} (2005a)]{BNFA05a}{\sc Brown, E.,  Nikolaenko, A., Funfschilling, D., \& Ahlers, G.} 2005a Heat transport in turbulent Rayleigh-B\'enard convection: Effect of finite top- and bottom-plate conductivity. {\em Phys. Fluids}, in print.

\bibitem[Brown {\em et~al.\/} (2005b)]{BNA05b}{\sc Brown, E., Nikolaenko, A., \& Ahlers, G.} 2005b Orientation changes of the large-scale circulation in turbulent Rayleigh-B{\'e}nard convection. {\em Phys. Rev. Lett.}, submitted.

\bibitem[Brown {\em et~al.\/} (2005c)]{BNA05c}{\sc Brown, E., Nikolaenko, D., \& Ahlers, G.} 2005c , Effect of Earth's Coriolis force on the large-scale circulation in turbulent Rayleigh-B{\'e}nard convection, unpublished.

\bibitem[Castaing {\em et~al.\/}(1998)]{CGHKLTWZZ89}{\sc Castaing, B., G. Gunaratne, G., Heslot, F., Kadanoff, L., Libchaber, L., Thomae, S., Wu, X.-Z., Zaleski, S., \&  Zanetti, G.} 1998 Scaling of hard thermal turbulence in {{Rayleigh-B\'enard}} convection. {\em J. Fluid Mech.} {\bf 204}, 1--30.

\bibitem[Chaumat {\em et~al.\/}(2002)]{CCC02}{\sc Chaumat, S., Castaing, B., \& Chill\`a, F.} 2002 Rayleigh-B\'enard cells: influence of the plates properties {\em Advances in Turbulence IX, Proceedings of the Ninth European Turbulence Conference}, edited by I.P. Castro and P.E. Hancock (CIMNE, Barcelona) .

\bibitem[Chill\`a {\em et~al.\/}(2004)]{CRCC04}{\sc Chill\`a, F., Rastello, M., Chaumat, S., \& Castaing, B.} 2004b Long relaxation times and tilt sensitivity in Rayleigh-B\'enard turbulence. {\em Euro. Phys. J. B} {\bf 40}, 223--227.

\bibitem[Ciliberto {\em et~al.\/}(1997)]{CCL96}{\sc Ciliberto, S., Cioni, S., \& Laroche, C,} 1996 Large-scale flow properties of turbulent thermal convection. {\em Phys. Rev. E} {\bf 54}, R5901--R5904.

\bibitem[Cioni {\em et~al.\/}(1997)]{CCS97}{\sc Cioni, S., Ciliberto, S., \& Sommeria, J.} 1997 Strongly turblent  Rayleigh-B\'enard convection in mercury: comparison with results at moderate Prandtl number. {\em J. Fluid Mech.} {\bf 335}, 111--140.

\bibitem[Funfschilling {\em et~al.\/} (2005)]{FBNA05}{\sc Funfschilling, D., Brown, E., Nikolaenko, A., \& Ahlers, G.} 2005 Heat transport by turbulent Rayleigh-B\'enard Convection in cylindrical samples with aspect ratio one and larger. {\em J. Fluid Mech.}, in print.

\bibitem[Grossmann \& Lohse (2001)]{GL01}{\sc Grossmann, S. \& Lohse, D.} 2001 Thermal convection for large Prandtl number. {\em Phys. Rev. Lett.} {\bf 86}, 3317--3319.

\bibitem[Grossmann \& Lohse (2002)]{GL02}{\sc Grossmann, S. \& Lohse, D.} 2002 Prandtl and Rayleigh number dependence of the Reynolds number in turbulent thermal convection. {\em Phys. Rev. E} {\bf 66}, 016305 1--6.

\bibitem[Hansen {\em et~al.\/}(1991)]{HYK91} {\sc Hansen, U., Yuen, D. A.,  \& Kroening, S.E.} 1991 Mass and heat transport in strongly time-dependent thermal convection at infinite Prandtl number. {\em Geophys. Astrophys. Fluid Dynamics} {\bf 63}, 67--89.

\bibitem[Kadanoff(2001)]{Ka01} {\sc Kadanoff, L.~P.} 2001 Turbulent heat flow: Structures and scaling. {\em Phys. Today} {\bf 54} (8), 34--39.

\bibitem[Kraichnan(1962)]{Kr62} {\sc Kraichnan, R.} 1962 Turbulent thermal convection at arbitrary Prandtl number. {\em Phys. Fluids} {\bf 5}, 1374--1389.

\bibitem[Krishnamurty \& Howard(1981)]{KH81}{\sc Krishnamurty, R. \& Howard, L.~N.}1981 Large-scale flow generation in turbulent convection. {\em Proc. Nat. Acad. Sci. USA} {\bf 78}, 1981--1985.


\bibitem[Lam {\em et~al.\/}(2002)]{LSZX02} {\sc Lam, S., Shang, X.-D.,  \& Xia, K.-Q.} 2002 Prandtl number dependence of the viscous boundary layer and the Reynolds numbers in Rayleigh-B\'enard convection. {\em Phys. Rev. E} {\bf 65}, 066306 1--8.

\bibitem[Lui \& Xia(1998)]{LX98}{\sc Lui, S.-L. \& Xia, K.-Q.}1998 Spatial structure of the thermal boundary layer in turbulent convection. {\em Phys. Rev. E} {\bf 57}, 5494--5503.

\bibitem[Niemela {\em et~al.\/} (2001)]{NSSD01}{\sc Niemela, J., Skrbek, L., Sreenivasan, K., \& Donnelly, R.} 2001 The wind in confined thermal Convection. {\em J. Fluid Mech.} {\bf 449}, 169--178.

\bibitem[Nikolaenko {\em et~al.\/} (2005)]{NBFA05}{\sc Nikolaenko, A. , Brown, E., Funfschilling, D., \& Ahlers, G.} 2005 Heat transport by turbulent Rayleigh-B\'enard Convection in cylindrical cells with aspect ratio one and less. {\em J. Fluid Mech.} {\bf 523}, 251--260.


\bibitem[Roche {\em et~al.\/} (2004)]{RCCH04}{\sc Roche, P.-E., Castaing, B., Chabaud, B., \& H\'ebral, B.} 2004 Heat transfer in turbulent Rayleigh-B\'enard convection below the ultimate regime. {\em J. Low Temp. Phys.} {\bf 134}, 1011 -- 1042.


\bibitem[Shraiman and Siggia (1990)]{SS90}  {\sc B. I. Shraiman \& E. D. Siggia} 1990 Heat transport in high-{{Rayleigh}} number convection. {\em Phys. Rev. A} {\bf 42}, 3650--3653.

\bibitem[Siggia(1994)]{Si94} {\sc Siggia, E.~D.} 1994 High Rayleigh number convection. {\em Annu. Rev. Fluid Mech.} {\bf 26}, 137--168.

\bibitem[Qiu and Tong (2001a)]{QT01a}{X.-L. Qiu \& P. Tong} 2001 Large-scale velocity structures in turbulent thermal convection. {\em Phys. Rev. E} {\bf 64}, 036304 1--13.

\bibitem[Qiu and Tong (2001b)]{QT01b}{X.-L. Qiu \& P. Tong} 2001 Onset of coherent oscillations in turbulent Rayleigh-B\'enard convection. {\em Phys. Rev. Lett.} {\bf 87}, 094501 1--4.

\bibitem[Qiu and Tong (2002)]{QT02}{X.-L. Qiu \& P. Tong} 2002 Temperature oscillations in turbulent Rayleigh-B\'enard convection. {\em Phys. Rev. E} {\bf 66}, 026208 1--11.

\bibitem[Sreenivasan {\em et~al.\/} (2002)]{SBN02}{\sc Sreenivasan, K., Bershadskii, A., \& Niemela, J.} 2002 Mean wind and its reversal in thermal convection. {\em Phys. Rev. E} {\bf 65}, 056306 1--11.

\bibitem[Stringano \& Verzicco(2005)]{SV05} {\sc Stringano, G. \& Verzicco, R.} 2005 Mean flow structure in thermal convection in a cylindrical cell of aspect-ratio one half. {\em J. Fluid Mech.}, submitted.


\bibitem[Sun {\em et~al.\/} (2005)]{SXX05}{\sc Sun, C., Xi, H.-D., \& Xia, K.-Q.} 2005 Azimuthal symmetry, flow dynamics, and heat flux in turbulent thermal convection in a cylinder with aspect ratio one-half. {\em Phys. Rev. Lett}, submitted.


\bibitem[Verzicco \& Camussi(2003)]{VC03} {\sc Verzicco, R. \& Camussi, R.} 2003 Numerical experiments on strongly turbulent thermal convection in a slender cylindrical cell. {\em J. Fluid Mech.} {\bf 477}, 19--49.

\bibitem[Verzicco(2004)]{Ve04} {\sc Verzicco, R.} 2004 Effects of non-perfect thermal sources in turbulent thermal convection. {\em Phys. Fluids} {\bf 16}, 1965--1979.

\bibitem[Wu \& Libchaber(1991)]{WL91} {\sc Wu, X.-Z. \& Libchaber, A.} 1991 Non-{{Boussinesq}} effects in free thermal convection. {\em Phys. Rev. A} {\bf 43}, 2833--2839.

\bibitem[Zhang {\it et al.}(1997)]{ZCL97} {\sc Zhang, J., Childress, S., \& Libchaber, A.} 1997 Non-Boussinesq effect: Thermal convection with broken symmetry. {\em Phys. Fluids} {\bf 9}, 1034-1042.


\end{thebibliography}
\end{document}